\documentclass[sn-mathphys,
amsmath,amssymb,amsgen,amssymb, amsbsy, amsfonts
]{revtex4-1}

\usepackage{graphicx}
\usepackage{dcolumn}
\usepackage{bm}

\usepackage[utf8]{inputenc}
\usepackage[T1]{fontenc}
\usepackage{mathptmx}
\usepackage{etoolbox}
\usepackage{orcidlink}

\def\Rvsd#1{\textcolor{blue}{#1}}
\def\Rvsd#1{#1}

\makeatletter
\def\@email#1#2{%
 \endgroup
 \patchcmd{\titleblock@produce}
  {\frontmatter@RRAPformat}
  {\frontmatter@RRAPformat{\produce@RRAP{*#1\href{mailto:#2}{#2}}}\frontmatter@RRAPformat}
  {}{}
}%
\makeatother
\begin{document}


\title[Diffusion in the presence of a chiral topological defect
]{Diffusion in the presence of a chiral topological defect}
\author{A. Manapany}
\affiliation{Laboratoire de Physique et Chimie Th\'eoriques, Universit\'e de Lorraine, 54506 Vandoeuvre-les-Nancy Cedex, France.}
\affiliation{${\mathbb L}^4$ Collaboration \& Doctoral College for the
Statistical Physics of Complex Systems,
Leipzig-Lorraine-Lviv-Coventry}
\author{L. Moueddene} 
\affiliation{Laboratoire de Physique et Chimie Th\'eoriques, Universit\'e de Lorraine, 54506 Vandoeuvre-les-Nancy Cedex, France.}
\affiliation{${\mathbb L}^4$ Collaboration \& Doctoral College for the
Statistical Physics of Complex Systems,
Leipzig-Lorraine-Lviv-Coventry}
\author{B. Berche\orcidlink{0000-0002-4254-807X}}
\affiliation{Laboratoire de Physique et Chimie Th\'eoriques, Universit\'e de Lorraine, 54506 Vandoeuvre-les-Nancy Cedex, France.}
\affiliation{${\mathbb L}^4$ Collaboration \& Doctoral College for the
Statistical Physics of Complex Systems,
Leipzig-Lorraine-Lviv-Coventry}
\author{S. Fumeron} 
\affiliation{Laboratoire de Physique et Chimie Th\'eoriques, Universit\'e de Lorraine, 54506 Vandoeuvre-les-Nancy Cedex, France.}

\date{\today}

\begin{abstract}
We study the diffusion processes of a real scalar field in the presence of the distorsion field induced by a chiral topological defect. The defect modifies the usual Euclidean background geometry into a non-diagonal Riemann-Cartan geometry characterized by a singular torsion field. The new form of the diffusion equation is established and the scalar field distribution in the vicinity of the defect is investigated numerically. Results show a high sensitivity to the boundary conditions. In the transient regime, we find that the defect vorticity generates an angular momentum associated to the diffusion flow and we discuss its main properties.
\end{abstract}

\maketitle

\section{Introduction}

\Rvsd{In this paper, we propose a model describing the diffusion of a non relativistic scalar field in the vicinity of a chiral topological defect. The physical context can be that of material sciences and condensed matter physics. In particular, it is well-established that ductility is determined by dislocation mobility, and more especially that of screw dislocations (as opposed to edge dislocations, more easily pinned by obstacles they can climb by vacancy diffusion). Screw dislocations are also of interest in nondestructive-testing \cite{Schmerr1998}, management of fatigue fracture in materials \cite{Suresh1998} or heat flux design \cite{Fumeron2013}. From the microscopic standpoint, topological defects are likely to affect phonons transport, which, at a macroscopic scale, modifies heat conduction. Usually, elasticity theory incorporates defects at the expansive cost of a complicated set of boundary conditions. That is why a more fundamental theory has been developed, in which defects are described in terms of differential geometry: the strain field in the material is then described at large distances in the continuum limit as an effective metric\cite{Lazar2009}.} 

\Rvsd{Another context of interest for chiral topological defects is cosmology where,} using the correspondence between Schr\"odinger equation and a diffusion equation in imaginary time, the geometric approach developed here is able to describe non relativistic spinless particles ``trapped'' in the metric field of a chiral cosmic string\cite{GaltsovLetelllier1993,BezerraEtal2002}. The link between these two kinds of defects, namely \Rvsd{compact} screw dislocations and cosmic strings, and a review of the relevant metrics can be found in Ref.\cite{Puntigam1997} To make things concrete, we will mainly use the vocabulary of the first situation and call ``temperature'' the scalar field under consideration. In the same vein, the dislocation geometry will be established in the context of classical elasticity, but all the results remain directly transposable to the cosmological counterpart.

In section II, we will take a detour through analogue gravity and show that the screw dislocation vicinity can elegantly be described by an effective Riemann-Cartan geometry \cite{Lazar2009,Lazar2010,Kleinert1990}. \Rvsd{Such situation involves torsion and is likely to reveal new phenomena when considering transport.} The diffusion equation will be formulated along Smerlak's approach\cite{SmerlakPRE2012} as a sequence of Markovian processes ruled by a parabolic equation. In section III, the system will be solved numerically for various boundary conditions: the main outcome is that a screw dislocation imprints torsion onto the fluxes of the scalar field. A parametric study completes this part, emphasizing the role of the Burgers vector on diffusion. We evaluate in Section IV the  \Rvsd{heat flow angular momentum resulting from the torsion} and we find that the transient regime is characterized in this geometry by a non zero value which depends on time, on the geometric parameters and on the initial conditions. A discussion on how the defect can be used to functionalize a material concludes this article.

\section{The geometric model}

\subsection{Analogy between elasticity and 3D-gravity}

The geometric theory of defects originates in the pioneering works by Bilby \cite{Bilby1955} and Kr\"oner \cite{Kroner1958} in the 1950s and ever since, this approach has been extensively developped \cite{Kleman1980,Katanaev2005}. In essence, the geometric theory of elasticity represents a body submitted to internal stresses as a three dimensional continuum that can assume two possible states \cite{landau7}:
\begin{enumerate}
	\item[--] \textit{the ground (or initial) state} corresponds to the situation where the material is not deformed and as it is, it is maximally symmetric (invariance under translations and rotations). In such state, the medium is associated to a flat manifold equipped with an orthonormal Cartesian coordinate system $x^i$ ($i=1,2,3$). Lengths squared are given by the usual form of Pythagoras theorem:
	\begin{equation}
	dl_{gs}^2=\delta_{ij}dx^i dx^j=dx^2+dy^2+dz^2
	\end{equation}
	where $\delta_{ij}$ is the Kronecker symbol and the subscript $_{gs}$ stands for ground state.
	\item[--] \textit{the deformed (or final) state}: anytime an elasto-plastic deformation occurs, some of the symmetries mentioned hereinbefore are broken and the medium changes in size and/or in shape: as the points are moving with respect to each other, distances between them are changing as well and so is the associated metric $g_{ij}$. 
\end{enumerate}
	
The expression of $g_{ij}$ can be obtained in the case of linear elasticity from the displacement field $u^i(\textbf{x})$. For a point $M$ originally at position $x^i$ in the ground state, that same point is shifted to the position $y^i=x^i+u^i(\textbf{x})$ in the deformed state. The distance between any pair of points after deformation then writes:
	\begin{eqnarray}
	dl^2&=&\delta_{ij}dy^i dy^j=\left(\delta_{ij}+\partial_i u_j +\partial_j u_i +\delta_{kl}\partial_i u^k\partial_j u^l\right) dx^i dx^j
 \end{eqnarray}
Therefore, the elastic strain metric is obtained as
\begin{equation}
g_{ij}=\delta_{ij}+\partial_i u_j +\partial_j u_i +\delta_{kl}\partial_i u^k\partial_j u^l=\delta_{ij}+2\varepsilon_{ij} \label{analog-elasticity}
\end{equation}
where $\varepsilon_{ij}$ denotes the strain tensor. 

This short overview of the continuum theory of defective crystals calls for two remarks. First, a crystal consists in a discrete lattice of atoms and modeling it as a continuous medium is not self-explanatory. Rigorously, the continuum limit for elasticity should come as a coarse grained approximation of molecular dynamics and it should fail at the atomic scale (for attempts to push these limits down, see for instance Ref.~\onlinecite{Charlotte2012}). This is obvious when examining the concept of elastic strain, which is defined as an average value over a domain containing a sufficiently large number of unit cells, but that remains sufficiently smaller than the bulk size. The continuum approach is thus relevant when the characteristic length scale is large enough compared to interatomic distances: as noticed by Davini \cite{Davini1986}, this holds for instance for X-rays soundings, which provide information at scales where defects are smoothed out. 

The second remark is related to the status of (\ref{analog-elasticity}), which obviously possesses non-vanishing curvature and/or torsion, as in three-dimensional gravity. Yet, the deformed solid actually lives in a three-dimensional Euclidean space, which means that the geometry is flat. How is it possible to reconcile these two standpoints? As suggested by de Wit\cite{DeWit1981}, the state described by (\ref{analog-elasticity}) cannot exist in the flat space, but only in an imaginary space where the solid is relaxed: $g_{ij}$ comes from the projection of this imaginary space onto the physical flat space, in a similar way as a stereographic  projection transfers the geometric properties on a 2-sphere (the Earth, with its meridians and parallels) onto a flat plane while deforming them (Wulff net). It turns out that the geometric description of solids thus requires two metrics: Firstly the physical flat metric, $\delta_{ij}$, will be used to perform operations on tensors such as raising/lowering indices...  and secondly the effective metric $g_{ij}$, which contains the elastic information, will be used to determine the kinematics of low energy perturbations (geodesics, first integrals...). 

\subsection{Geometry of the screw dislocation}

\begin{figure}[h]
\includegraphics[width=9cm]{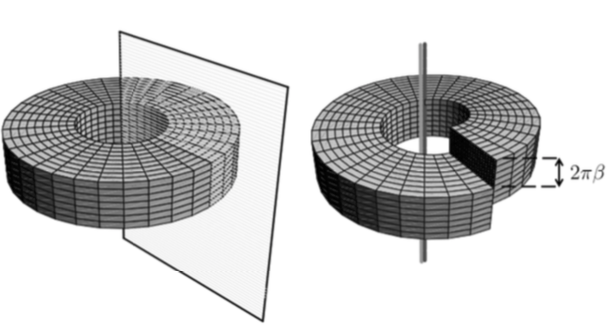} 
\caption{Volterra cut-and-weld process for a screw dislocation along z (taken from \cite{Puntigam1997} and adapted by the authors)}\label{Volterra1}
\end{figure}

Screw dislocations are line defects related to shear stress that are very common in crystalline solids. They are associated to a breaking of the translational symmetry group and can be easily pictured by a Volterra cut-and-weld process (see Fig. \ref{Volterra1}). In this example, the magnitude and the direction of the crystal lattice distortion resulting from a screw dislocation of axis $z$ is encompassed in the Burgers vector $\mathbf{b}=b\:\mathbf{e_z}$. In real crystals, $b$ is a small parameter that is usually of the same order as the unit cell length.
The displacement field at equilibrium obeys the two relations:
\begin{eqnarray}
\Delta u_z&=&0 \\
\oint_{\mathcal{C}}du_i&=&-b_i
\end{eqnarray}
Solutions of these equations were found in 1907 by Volterra to be \cite{landau7}:
\begin{equation}
u_z=\frac{b}{2\pi}\theta=\beta\theta
\end{equation}
Computing the components of the strain tensor leads straightforwardly to
\begin{eqnarray}
dl^2=g_{ij}dx^i dx^j=dr^2+r^2d\theta^2+\left(\beta d\theta+dz\right)^2 \label{SD-metric} \label{metric}
\end{eqnarray}
\Rvsd{where the Einstein summation convention of repeated indices was used.} This line element can be understood as follows: by performing clockwise a complete turn around the axis, one moves up by one unit of Burgers vector $\textbf b$. This can be summed up by the condition that $\theta\rightarrow\theta+2\pi$ leads to $z\rightarrow z+b$. 

The torsion tensor \cite{carroll,Hammond02} has only one non-vanishing component given by
\begin{equation}
T^z_{\;r\theta}=2\pi\beta \delta^2\left(r\right) \label{torsion-tensor}
\end{equation}
whereas the curvature tensor is identically zero. Hence, a screw dislocation must be described in terms of a Riemann-Cartan manifold, for which torsion is only located on the dislocation axis and vanishes everywhere else. Note that the infinitely thin approximation for the dislocation line is a natural outcome of the above discussion on length scales but does not presume anything about the relevance of a core structure. 

One may naturally wonder what really happens on the defect axis and if the zero-width model can be refined. In hard matter, the high-strained core region is of atomic width and strictly speaking, the continuum approximation breaks down: classical elasticity gets burdened by singularities (as illustrated by the delta functions in Eq. (\ref{torsion-tensor})) and it must be dropped in favor of atomistic calculations (empiric potentials or ab-initio simulations). Yet, for dislocations, all the singularities can be removed by considering an extended tubular core with an isotropically-distributed Burgers vector about every point on the defect axis \cite{Cai2006}. \Rvsd{A limitation which is not studied in this paper is linked to the amplitude of the Burgers vector. Indeed, ought to Frank's energy criterion, dislocations in face-centered-cubic metals generally split into two Shockley when the Burgers vector becomes too large. This effect, promoted by a weakening of the total energy associated to the presence of the two partial defect, is not explicitly taken into account since the parameter $\beta$ which measures the Burgers vector below is not limited. The values considered are rather chosen for the sake of clarity of the illustrations but the observed phenomena display continuity wrt these values and consolidate our main conclusions.}


\subsection{Diffusion processes in a non-euclidean geometry}

Generally speaking, diffusion of a passive scalar (for instance the temperature field) can be seen as a collection of Markov processes obeying the stochastic Fokker-Planck equation. In the case of Brownian motion, the Fokker-Planck equation reduces to the well-known parabolic heat equation \cite{Pavliotis2014}. When considering diffusion processes in the presence of a non-Euclidean space, the problem is addressed by replacing the Laplace operator with the Laplace-Beltrami operator $\Delta_{lb}$ \cite{SmerlakPRE2012}: 
\begin{equation}
\frac{\partial T}{\partial t}=D \Delta_{lb} T \label{Eckart}
\end{equation}
Here, $D$ is the diffusivity and its value depends on the material, and the Laplace-Beltrami operator writes as
\begin{eqnarray}
   \Delta_{lb}&=&\frac{1}{\sqrt{g}}\frac{\partial}{\partial x^i}\left(\sqrt{g}g^{ij}\frac{\partial}{\partial x^j} \right) \label{LB1}
\end{eqnarray}
for a background geometry described by the static metric $g_{ij}$. Equation (\ref{Eckart}) is a particular case of the Eckart equation \cite{Eckart1940}, first introduced in 1940 to study the irreversible thermodynamics of relativistic fluids and later refined by Landau and Lifshitz in 1959 \cite{Peitz1998}.

For the screw dislocation, \Rvsd{the line element (\ref{metric}) is equivalent to the effective metric \cite{Fumeron2013}}
\begin{equation}
(g_{ij}) = 
\begin{pmatrix}
 1 & 0 & 0 \\
 0 & r^2+\beta^2 & \beta \\
 0 & \beta & 1
\end{pmatrix}
\end{equation}
and therefore, the generalized diffusion equation writes as:
\begin{eqnarray}
   \frac{\partial T}{\partial t}&=&D\left[\frac{\partial^2 T}{\partial r^2}+\frac{1}{r}\frac{\partial T}{\partial r} + \frac{1}{ r^2}\frac{\partial^2 T}{\partial \theta^2}+
   \left(\frac{\beta^2}{r^2}+1\right)\frac{\partial^2 T}{\partial z^2}\right. \nonumber \\
   &&-\left.\frac{2\beta}{r^2}\frac{\partial^2 T}{\partial \theta \partial z} \right]\label{PDE}
\end{eqnarray}

As a comment, we should mention that  the equivalence between Schr\"odinger's equation and the diffusion equation has been extensively investigated and it turns out to be a fruitful way of getting analytical solutions (see for instance Ref. \onlinecite{Nagawa1993}). In the presence of a screw dislocation, the dynamics of a free particle obeys:
\begin{equation}
i\hbar\frac{\partial T}{\partial t}=-\frac{\hbar^2}{2m}\Delta_{lb} T \label{schrod}
\end{equation}
Performing Wick's rotation $\tilde{t}=(i\hbar/2m) t$ allows to connect the solutions provided in Ref. \onlinecite{Bausch1999, Furtado2008} to our problem. 
The next sections are dedicated to numerical resolutions of (\ref{PDE}) in order to identify the peculiarities of diffusion processes when coupled to a chiral line defect.

\section{Results and discussion}

The solution $T(t,r,\theta,z)$ of the diffusion equation \ref{Eckart} was computed from Mathematica 12. To preserve the symmetries associated to the metric  (\ref{metric}), the domain of computation is chosen to be a cylinder of radius $R_1$ and of height $h$. To disregard any effect connected to the core region, the central part of the domain, a cylinder of radius $R_0$ (same height $h$), is removed. This ``inner boundary" is set at the cold normalized initial value $T_0=0$. The diffusion coefficient is taken equal to 1 (other values of the diffusion coefficient were also considered, but they did not add substantial phenomenolgy after a rescaling.).

The lateral walls  display adiabatic properties described by the boundary conditions (BC): 
\begin{equation}
    \mathbf{u}_r.\boldsymbol{\nabla}T|_{r=R_0}=\mathbf{u}_r.\boldsymbol{\nabla}T|_{r=R_1}=0
\end{equation}
Additionally, the various BC discussed hereafter will involve a hot normalized source $T_1=1$, and/or an incident heat flux (also normalized). As will be discussed hereafter, diffusion of heat flux couples both to the geometry (\ref{metric}) and to BC: as a matter of fact, depending on the symmetries of the BC, one may favor a preferentially upwards diffusion or a preferentially lateral diffusion in the presence of the screw dislocation. Here the terminology of ``upwards'' and ``lateral'' is approximate, because due to the chiral character of the metric, the coordinates $z$ and $\theta$ are coupled.
There is also a limiting behaviour, since the cylinder being of finite size, the stationary regime reached at $t\to\infty$ is anyway characterized by  a uniform temperature field.

\subsection{Preliminary observations: boundary conditions and geometrical couplings}
\begin{center}
\begin{figure}[t]
\includegraphics[width=8cm]{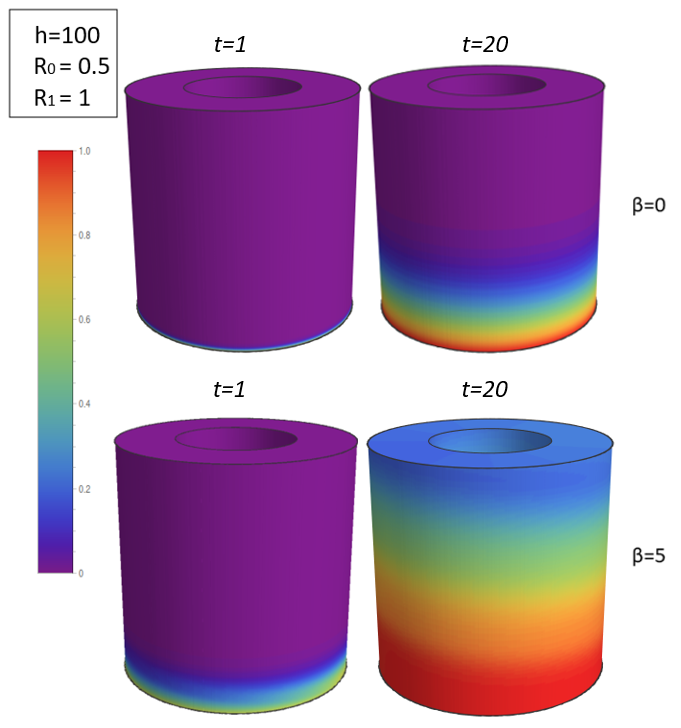}
\caption{Evolution of the diffusion profile for $\beta=0$ and $\beta=5$ at two time instants, $t=1$ and $t=20$ (arbitrary units). For better readability, the aspect ratio of the cylinder is not preserved.}\label{fig3}
\end{figure}
\end{center}
Let us start with a boundary condition consisting in a hot bottom disk:
\begin{eqnarray}
    T(0,r,\theta,z)&=&T_0=0 \ \ (\forall r,\theta,z;\ t=0)\\
    T(t,r,\theta,0)&=&T_1=1 \ \ (\forall r,\ \forall\theta,\ z=0;\ t>0)
\end{eqnarray}
Fig.~\ref{fig3} depicts the time evolution of the scalar field at two different instants, $t=1$ and $t=20$. All variables are dimensionless and $t=20$ corresponds to a situation where the medium is still far out of equilibrium. The screw dislocation appears to enhance the speed of the diffusion process. As a matter of fact, in absence of defect ($\beta=0$), the heated region barely reaches a third of the cylinder, whereas for $\beta=5$, the whole cylinder has been warmed up. A keen observation shows a  slightly tilted aspect of the diffusion front, as opposed to the Euclidean case where the front progresses upward uniformly. This is the mark of the screw character of the topological defect, and it turns out that with the present set of BC (heating of the full lower boundary), the coupling of the diffusion process with the defect geometry is weak. 

\subsection{Strong torsional regime} 

\begin{figure}[h]
\begin{center}\includegraphics[width=7cm]{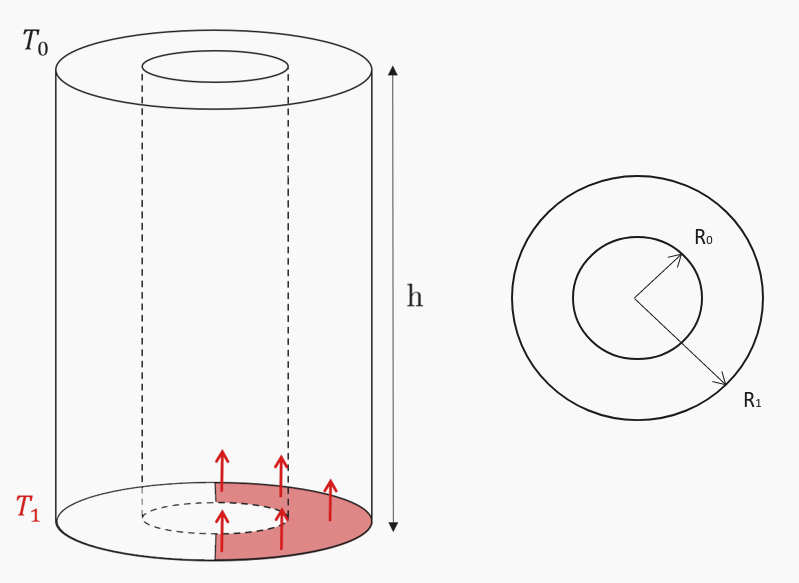} 
\caption{Boundary conditions enabling the upward diffusion  $T(t,r,\theta,0)=T_1,\ \forall r,\  0<\theta<\pi,\ \forall t>0$.}\label{fig4}
\end{center}
\end{figure}

We  consider now a new set of BC corresponding to a possibly stronger regime of coupling between heat transfer and the dislocation (see Fig.~\ref{fig4}). The lower boundary is heated up asymmetrically, only a half of it being set at the highest temperature  $T(t,r,\theta,0)=T_1$ for $0<\theta<\pi$ (other angular portions were also studied and essentially lead to similar results than those shown here). 
\begin{widetext}
\begin{center}
\begin{figure}[h]
\hfill\hbox to 0pt{\hss\includegraphics[width=13cm]{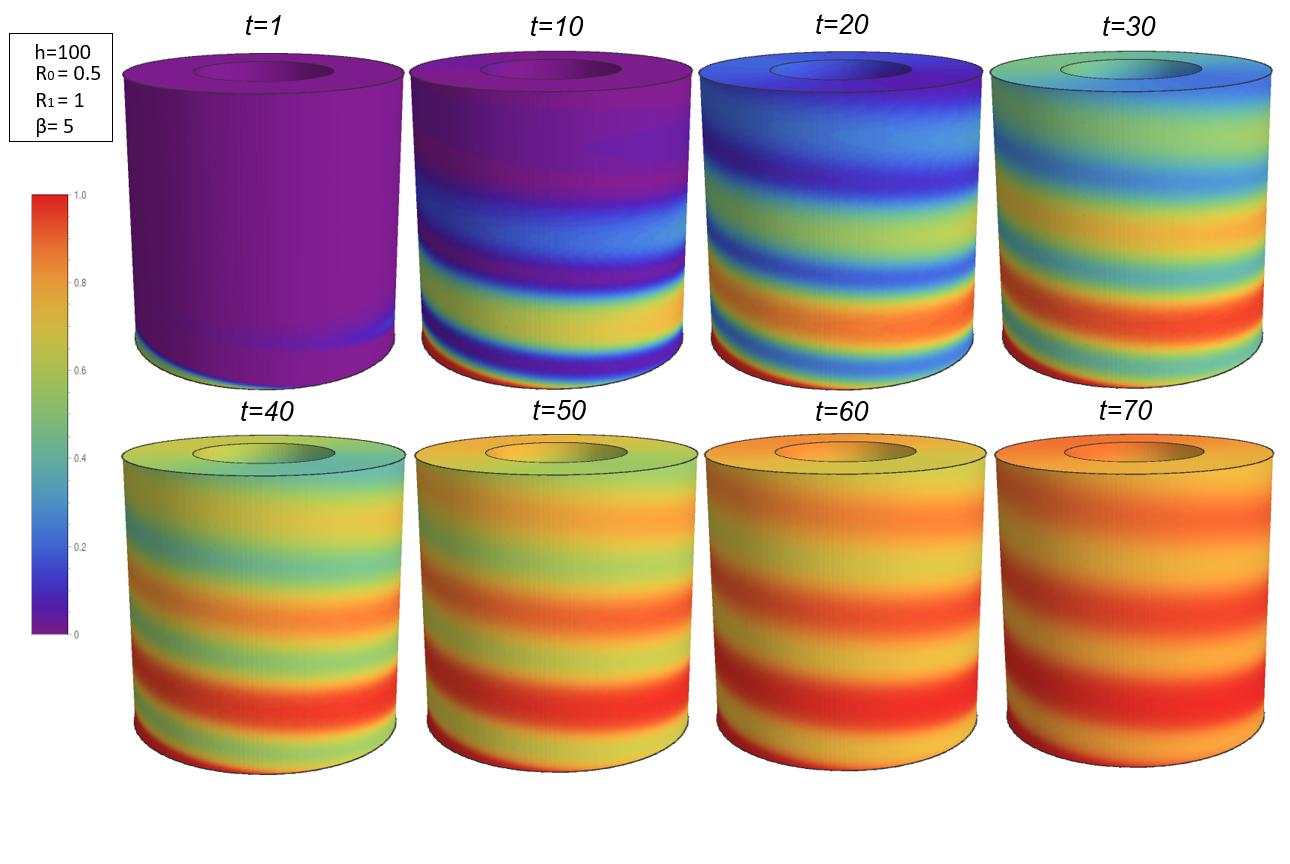}\hss}\hfill\null
\caption{Time evolution of the diffusion process for $\beta=5$. From top to bottom, left to right, time evolves between $t=1$ and $t=70$. The spiral shape of the temperature profile appears clearly. Comparison at $t=20$ can be done with the first uniform boundary condition in Fig.~\ref{fig3}.}\label{fig5}
\end{figure}
\end{center}
\end{widetext}
The results of the front diffusion at various times are obtained in Fig.~\ref{fig5}, where the time evolution of the diffusion process is presented at eight different instants from $t=1$ to $t=70$ for the same dislocation parameter $\beta=5$ than in the lower part of Fig.~\ref{fig3}. The asymmetric BC provides a contrasted helical diffusion pattern, which progresses faster than in the Euclidean case. Coiling of isotherms along the cylinder axis (clockwise here) is the direct manifestation of the chirality of the dislocation (the sign of $\beta$, which is positive in this case). A negative value of this parameter results in a counter-clockwise diffusion pattern, which is clearly emphasized on Fig.~\ref{fig6}. Additionally, it appears that the peaks of temperature are separated from each other by a fixed distance : as we will see in the last section, the pitch of the helix turns out to be equal to norm of Burgers vector $b=2\pi\beta$.
\begin{figure}[t]
\begin{center}
\includegraphics[width=8cm]{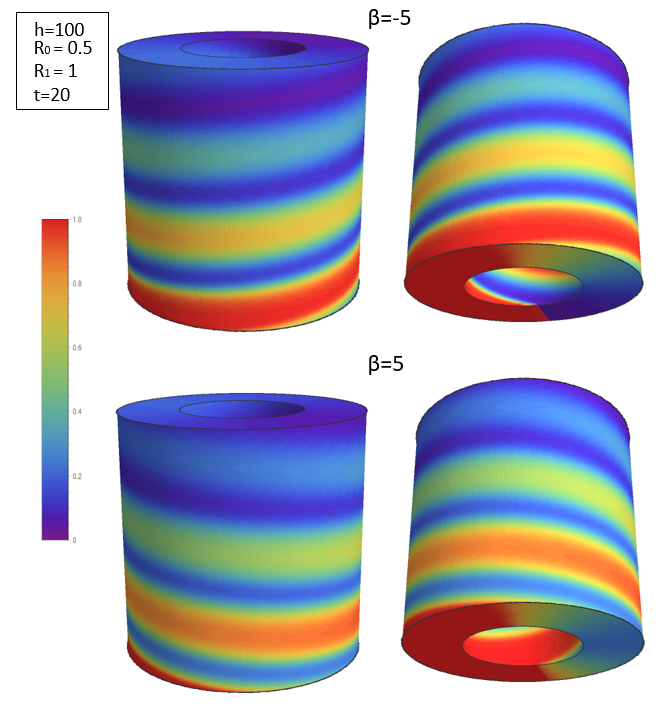} 
\caption{Chiral character of the diffusion pattern stemming from the sign of the dislocation parameter $\beta$. The temperature profile is shown at $t=20$.}\label{fig6}
\end{center}
\end{figure}

\begin{figure}[h]
\begin{center}
\includegraphics[width=7cm]{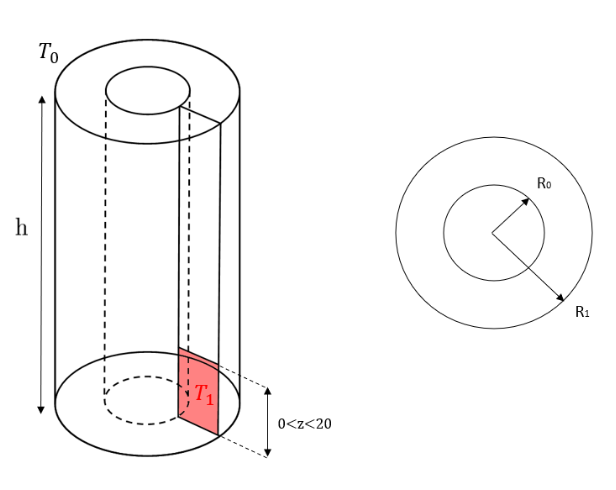} 
\caption{Boundary conditions enabling the  lateral diffusion.}\label{fig4bis}
\end{center}
\end{figure}

We now consider a second type of boundary conditions which will favor now a ``lateral" diffusion. To do so, we adopt the set of initial and boundary conditions depicted in Fig.~\ref{fig4bis}: the heating process only occurs on a fraction of a radial plane including the axis in the cylinder. Under these new conditions, we simulate the time evolution of the diffusion process with a dislocation parameter ranging from  $\beta=0$ to $\beta=14$. For both boundary conditions of
Figs.~\ref{fig4} (``upwards'' diffusion) and \ref{fig4bis} (``lateral'' diffusion), the temperature fields are shown at $t=30$ in Figs.~\ref{fig15} and \ref{fig15bis}. 
\begin{widetext}
\begin{center}
\begin{figure}[h]
\hfill\hbox to 0pt{\hss\includegraphics[width=13cm]{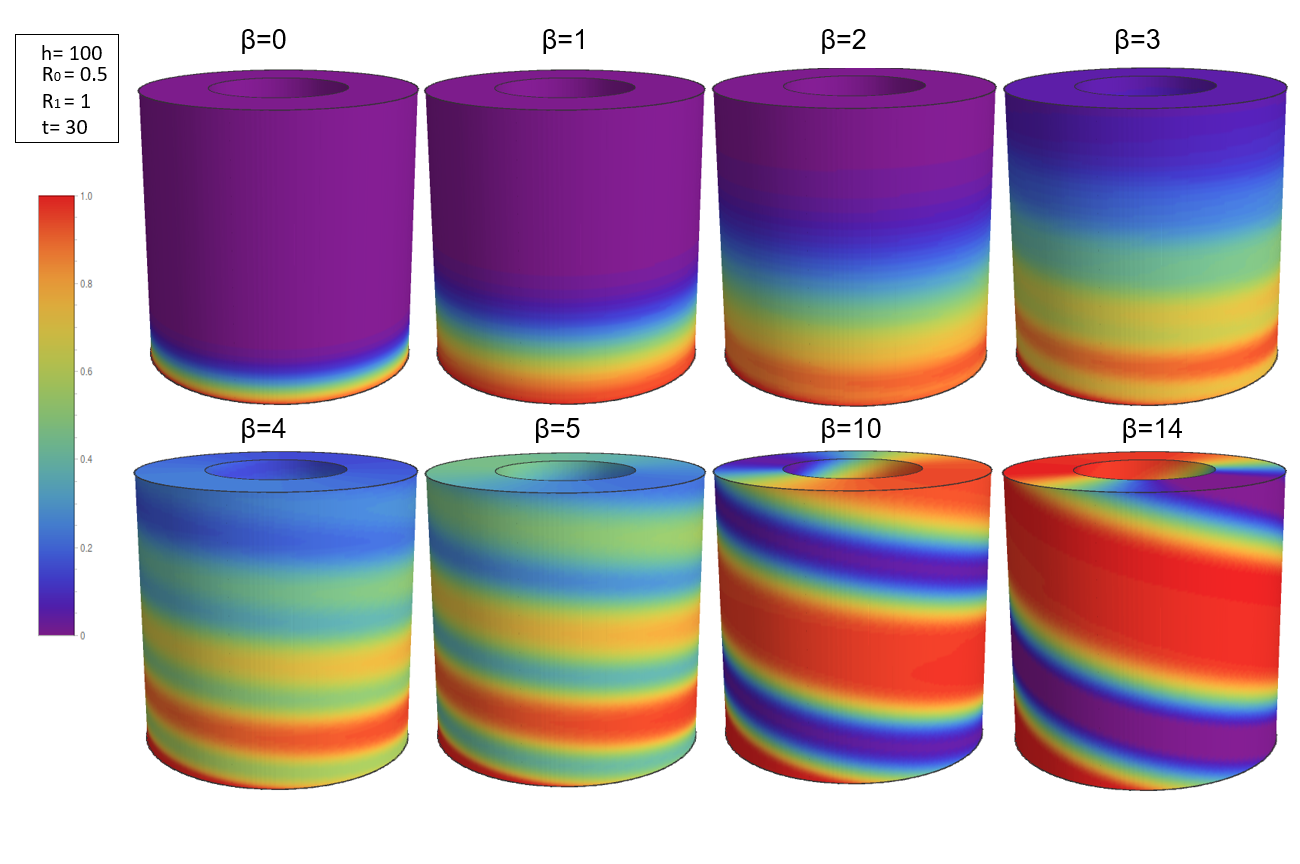} \hss}\hfill\null
\caption{Influence of the dislocation parameter $\beta$ on the dislocation parameter (upward diffusion). The profiles are shown at $t=30$ and the dislocation parameter is ranging from the euclidean case $\beta=0$ to $\beta=14$. The pitch of the helix displayed by the temperature profile increases with $\beta$.}\label{fig15}
\end{figure}
\end{center}
\end{widetext}

\begin{widetext}
\begin{center}
\begin{figure}[h]
\hfill\hbox to 0pt{\hss\includegraphics[width=13cm]{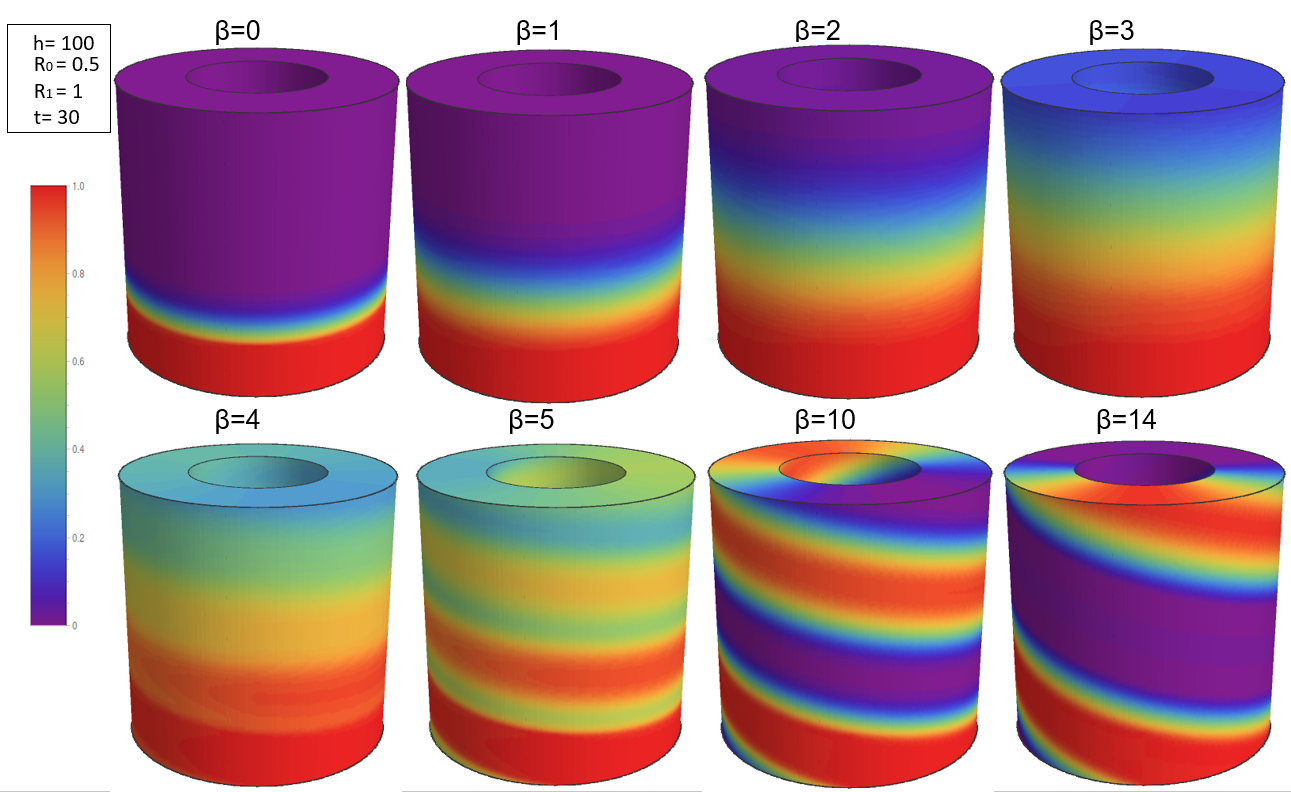} \hss}\hfill\null
\caption{Same as FiG.~\ref{fig15} for an angular diffusion.}\label{fig15bis}
\end{figure}
\end{center}
\end{widetext}

\subsection{Influence of the dislocation parameter $\beta$ on the diffusion process}

\begin{figure}[ht]
\begin{center}
\includegraphics[width=8cm]{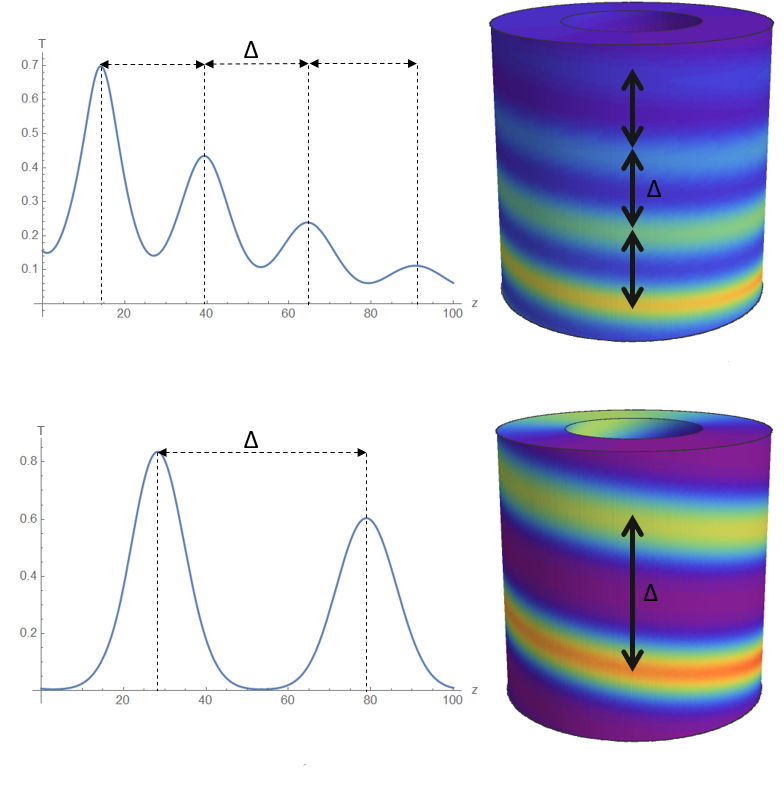} 
\caption{Measurement of the distance $\Delta(\beta)$ between each peak of temperature drawn at a fixed value of $t$,  $\theta$ and $r$: $T(t=30,r=0.5,\theta=\pi,z)$. The upper part is for $\beta=4$ and the lower part for $\beta=8$.}\label{fig8}
\end{center}
\end{figure}

We have noticed in Fig.~\ref{fig15}  that the influence of the dislocation parameter $\beta$ is directly highlighted by the distance between each peak of temperature shown on the various snapshots.
In order to get these distances, called $\Delta(\beta)$, we plot the solution $T(t,r,\theta,z)$ with respect to the height $z$, but with fixed values of the time $t$, the radial distance $r$ and the angle $\theta$. Further details is provided in Fig.~\ref{fig8}.
These distances between successive peaks of temperature are determined there.
The function $T(t=\hbox{const},r=\hbox{const},\theta=\hbox{const},z)$ displays an oscillating behavior which is quite in line with the diffusion profile shown underneath the plot. The peaks of temperature are each equidistant to one another, with said distance increasing with the value of the parameter $\beta$. One can also notice that the height of each peak steadily decreases with $z$. It simply shows that the diffusion process takes a certain time to reach the end of the cylinder (which would get a uniform temperature at infinite time).

\begin{figure}[ht]
\begin{center}
\vbox{\includegraphics[width=6cm]{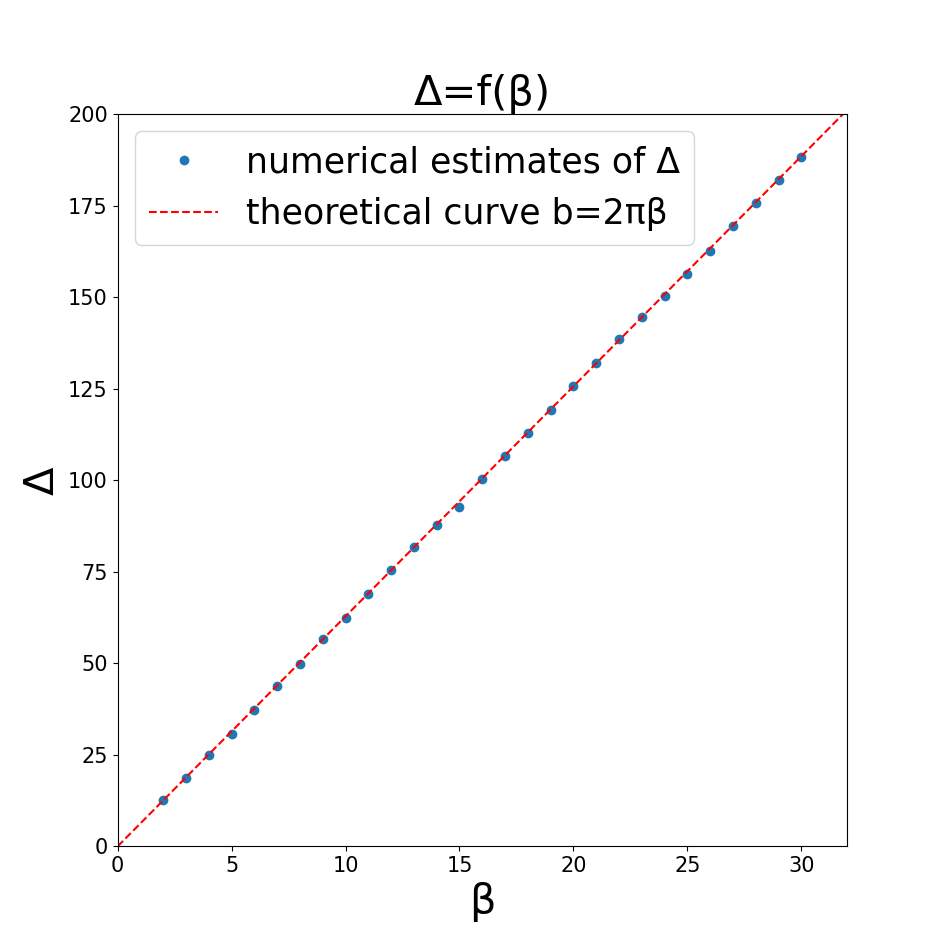}\includegraphics[width=6cm]{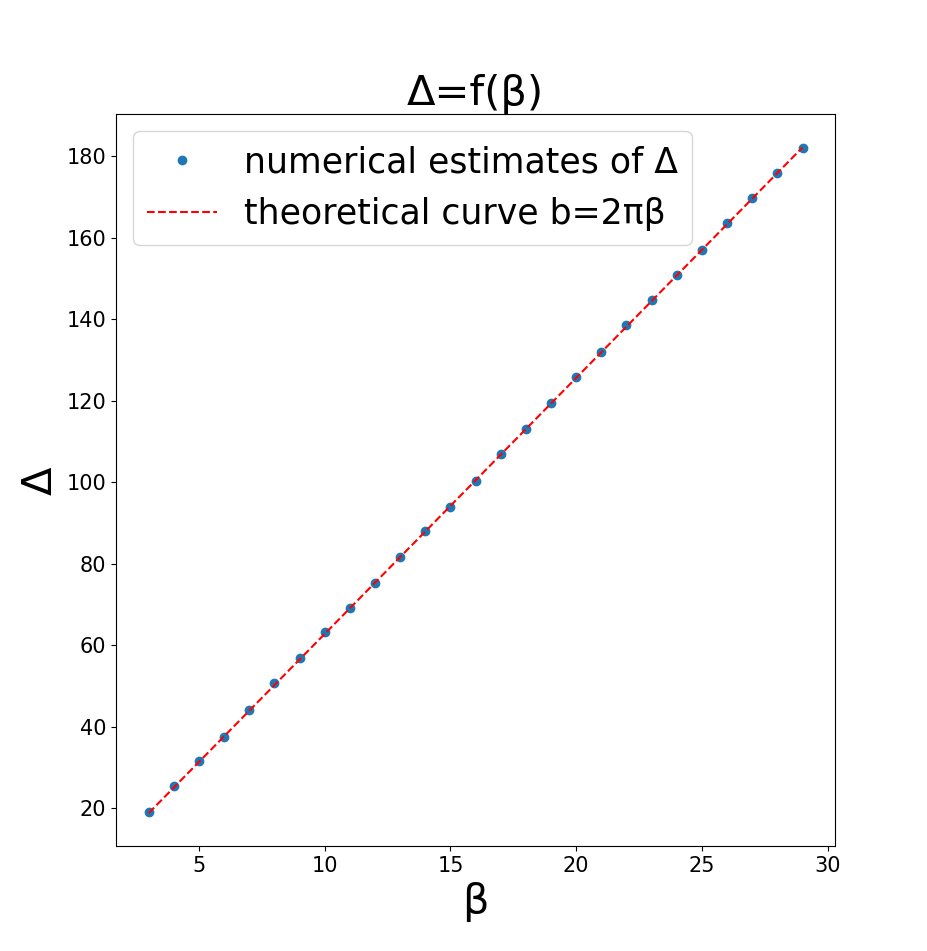}
}
\caption{Height differences between peaks of temperature with respect to the parameter $\beta$ (left: upward diffusion, right: angular diffusion).}\label{fig9}
\end{center}
\end{figure}

The distance $\Delta(\beta)$  increases with the value of the $\beta$ and the resulting values are reported in Fig.~\ref{fig9} versus $\beta$ for the two sets of boundary conditions and are compared
to the linear behavior $2\pi\beta$. The agreement is perfect: Evidently, the points recorded using the method described above falls almost exactly on the theoretical curve, thus validating the claim that the height difference between each spiral reporduce the geometry and corresponds to exactly $2\pi\beta$.

In the end, we can conclude than all the observed properties displayed for the upward diffusion process carry over to the rotational diffusion case. Whether it is the chiral aspect of the coiling motion, or the linear dependence of the height difference of each peaks, with respect to the parameter $\beta$, both diffusion processes behaves in a similar fashion. Indeed, such similarities are simply the result of the symmetrical role played by both the $z$ and $\theta$ variables displayed in the diffusion equation, when focusing solely on the cross-derivative term $-\frac{2\beta}{r^2}\partial_z\partial_{\theta}$.

\section{Angular momentum transfer}

 The heat flux is defined from Eckart's phenomenological constitutive relation, who derived it on the basis of thermodynamical arguments \cite{Eckart1940, Smerlak2011}. For the static single screw dislocation, its expresses as \cite{Fumeron2013IJTS} 
\begin{equation}
    q^{i}=-\kappa g^{ij} \frac{\partial T}{\partial x^i} 
\end{equation}
where $\kappa$ is the thermal conductivity of the bulk material. The connection with the orthogonal cartesian basis ($\hat{x}, \hat{y}, \hat{z}$) experienced by an external observer is provided by the co-tetrad fields~\cite{Ramond} 
 ${e^{\hat x}}_i=(\cos\theta,\ -r\sin\theta,\ 0)$, ${e^{\hat y}}_i=(\sin\theta,\ r\cos\theta,\ 0)$ and ${e^{\hat z}}_i=(0,\ \beta,\ 1)$.
Therefore, the components of the heat flux density are given by:
\begin{eqnarray}
    q^{\hat{x}}&=&-\kappa \cos\theta\frac{\partial T}{\partial r}+\frac{\kappa}{r} \sin\theta\left(\frac{\partial T}{\partial \theta}-\beta\frac{\partial T}{\partial z}\right) \\
    q^{\hat{y}}&=&-\kappa \sin\theta\frac{\partial T}{\partial r}-\frac{\kappa}{r} \cos\theta\left(\frac{\partial T}{\partial \theta}-\beta\frac{\partial T}{\partial z}\right) \\
    q^{\hat{z}}&=&-\kappa \frac{\partial T}{\partial z}
\end{eqnarray}
This entails the existence of a non-vanishing heat flux angular momentum density along the dislocation axis:
\begin{eqnarray}
    L_{\hat{z}}\propto \hat{x}q^{\hat{y}}-\hat{y}q^{\hat{x}}=-\kappa\left(\frac{\partial T}{\partial \theta}-\beta\frac{\partial T}{\partial z}\right)\label{eq_Lz}
\end{eqnarray}
In the absence of screw dislocation, $\beta=0$, and the mono-valuedness of the field $T$ prevents it to depend on $\theta$, then $ L_{\hat{z}}$ vanishes there as expected. 

\begin{figure}[ht]
\begin{center}
\includegraphics[width=8cm]{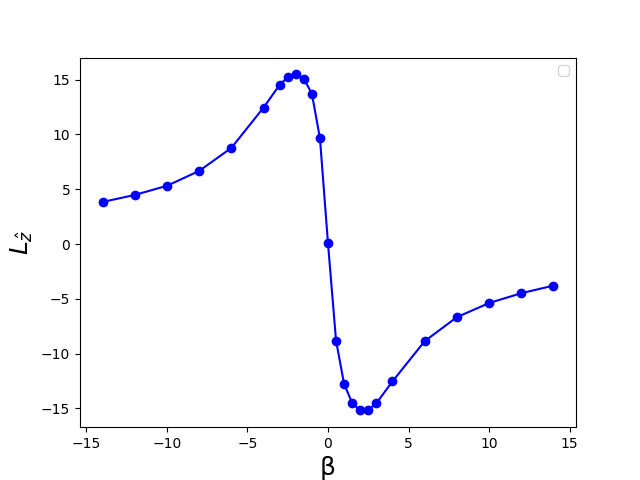}
\caption{Angular momentum $L_z$ as given by Eq.~(\ref{eq_Lz}) at time $t=100$. (upwards diffusion with fixed heat flux imposed on the half lower disk).}\label{fig11}
\end{center}
\end{figure}

\begin{figure}[ht]
\begin{center}
\includegraphics[width=8cm]{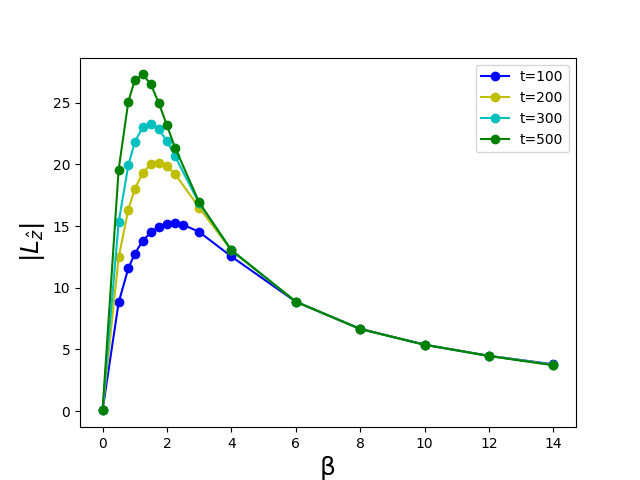}
\caption{Angular momentum $|L_z|$  at various values of time (upwards diffusion with fixed heat flux imposed on the half lower disk).}\label{fig12}
\end{center}
\end{figure}

The asymptotic stationary regime as we have already discussed is characterized by a uniform temperature field, and also corresponds to a vanishing angular momentum for arbitrary values of $\beta$.
At finite time on the other hand, there is a non-vanishing value of $L_z$ which displays a non monotonic behaviour (see Fig.~\ref{fig11}). The sign of $L_z$ is fixed (and opposite to it) by the sign of the dislocation paramater. The maximum value of $|L_z|$ depends on the details of the simulation, in particular on the time $t$. It shifts when $t$ increases to smaller values of $|\beta|$ in the case of a boundary condition imposed by a fixed heat flux (see Fig.~\ref{fig12}), but shifts the other way when the BC imposes a fixed value of the temperature.


\begin{figure}[ht]
\begin{center}
\vbox{\includegraphics[width=8cm]{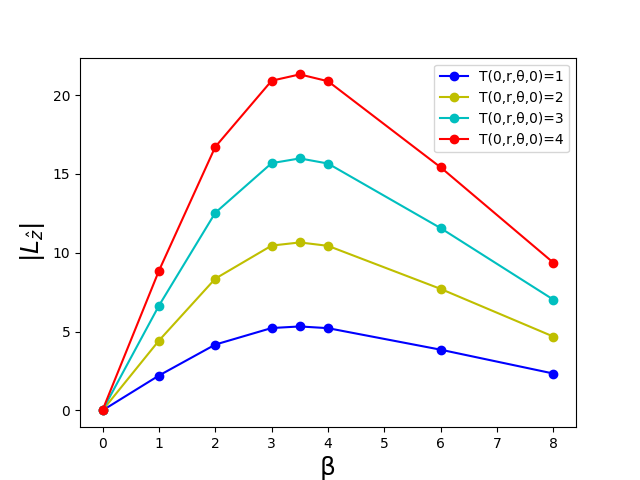}
\includegraphics[width=8cm]{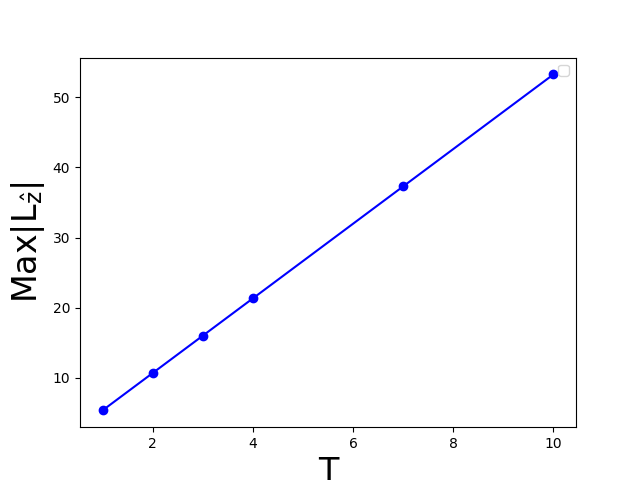}}
\caption{Left: Angular momentum $|L_z|$  for various values of the temperature imposed on the half lower disk). Right:  Linear dependence of the maximum with the value of $T$ at the BC.}\label{fig13}
\end{center}
\end{figure}

The height of the maximum of $|L_z|$ varies with the BCs and we find empirically a linear dependence of this maximum, at fixed time $t$, with the temperature imposed at the BC (see Fig~\ref{fig13}). The same appears to be  true when the BC imposes a fixed heat flux, there is a linear dependence of the maximum value of $|L_z|$ with the heat flux injected. 

\begin{figure}[ht]
\begin{center}
\includegraphics[width=8cm]{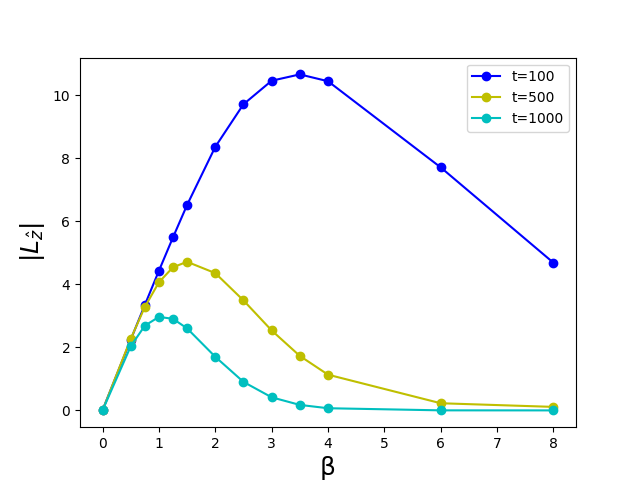}
\caption{Angular momentum $L_z$  at large values of time (upwards with fixed temperature imposed on the half lower disk).}\label{fig14}
\end{center}
\end{figure}

The angular momentum decays, as expected,  for large values of the diffusion time (see Fig.~\ref{fig14}), and would eventually vanish when $t\to\infty$.

\section{Concluding remarks}

In this paper, we modeled diffusion process in the non-Euclidean geometry generated by a chiral defect. Ought to the non vanishing torsion, the diffusion equation comprises a new term which couples the angle $\theta$ to the coordinate $z$ (in cylindrical coordinates). This term is responsible for the ``chiral'' character of the propagation of the scalar field. 

The presence of topological defects in condensed matter systems is usually a natural obstacle for the propagation of various physical signals (sound, heat, electronic conduction, etc). This drawback can be turned into an asset in the problem investigated here: indeed, using adequate boundary conditions, we showed that heat conduction can be exalted, since a channeling of the propagation of the warm front (resp. of the cold front) and a simultaneous increase in the speed of the diffusion process occur. Thermal design from screw dislocations is another illustration of the emerging field of defect engineering, which is now the object of a growing attention ought to its wide range of potential applications (heat transfer \cite{Fumeron2013IJTS}, spin transport \cite{Fumeronetal2017}...).

Another interesting outcome of our analysis is the possibility to use the heat flow to propose an analogue of the 
Einstein-de Haas experiment. This is a  famous experiments of the beginning of the \Rvsd{twentieth} century based on gyromagnetic 
phenomena~\cite{Richardson1908,Barnett1935,Frenkel79}. A cylinder made of a non magnetized ferromagnetic material,
suspended to a torsion wire, is subject to an external magnetic field 
along the cylinder's axis. The cylinder acquires a magnetization,  and at the
same time, the cylinder {starts a rotating motion}.
This effect became an efficient 
experimental method for the measurement of the gyromagnetic ratio of various 
materials and, for example, proved that magnetism of iron is essentially due
to the spin degrees of freedom.

This experiment is well known, and well understood from very fundamental 
principles, since it relies on the conservation of angular momentum.
When the material gets magnetized, the individual magnetic moments 
of the electrons point in a common direction, hence a total angular momentum
appears which has to be compensated by an opposite angular momentum carried
by the whole sample in order to conserve the initially vanishing value of the
total angular momentum of the sample. 

Here a similar situation can occur during the transient regime. The cylinder is initially at rest and has no net angular momentum. It is then submitted, say to an incident upwards heat flow, and as a consequence acquires an angular momentum
which has to be compensated by a global rotation of the cylinder. This appears as a kind of unexpected transfer of heat into work.

\begin{acknowledgments}
AM, LM and BB thank the ${\mathbb L}^4$ Collaboration, the  Doctoral College  
``Statistical Physics of Complex Systems -- 
Leipzig-Lorraine-Lviv-Coventry"  and the ``Universit\'e Franco-Allemande" for financial support.
\end{acknowledgments}

\section*{Author Contribution Statement}
All authors contributed equally to the paper.

\section*{Data Availability Statement}
Data sharing is not applicable to this article as no datasets were generated or analysed during the current study.

\end{document}